\documentclass[preprint,showpacs,preprintnumbers,amsmath,amssymb]{revtex4}
\usepackage{graphicx}
\usepackage{rotating}
\usepackage{epsf}
\usepackage{dcolumn}
\begin{document}
\preprint{preprint - thermoelectric group, iitp/iitb}
\title{Magnetism in La$_{0.7}$Sr$_{0.3}$Mn$_{1-x}$Co$_x$O$_3$ ($0 \leq x \leq 1$)}
\author{Ashutosh Kumar$^{1}$, Himanshu Sharma$^{2}$, C. V. Tomy$^{2}$, Ajay D. Thakur$^{1,\,3}$}
\affiliation{$^{1}$ Department of Physics, School of Basic Sciences, Indian Institute of Technology Patna, Patna 801 118, India \\
$^{2}$ Department of Physics, Indian Institute of Technology Bombay, Mumbai 400 076, India 
\\
$^{3}$ Center for Energy and Environment, Indian Institute of Technology Patna, Patna 801 118, India}
\date{\today}
\begin{abstract}
We study the structural and magnetic properties of La$_{0.7}$Sr$_{0.3}$Mn$_{1-x}$Co$_x$O$_3$ ($0 \leq x \leq 1$). Rietveld refinement of X-ray Diffraction (XRD) pattern suggests phase purity of the polycrystalline samples with R$\bar{3}$c space group. Interplay of Ferromagnetic (FM) and Antiferromagnetic (AFM) interaction upon Co substitution at Mn site in La$_{0.7}$Sr$_{0.3}$MnO$_3$ is evident from magnetic measurements. There is an optimal cobalt substitution at which the coercive field is maximum.
\end{abstract}
\maketitle

\section{Introduction}
Alkaline elements substituted rare earth manganites (RE$_{1-x}$A$_x$MnO$_3$), where RE=rare earth elements (La, Nd, Pr etc.) and A = alkaline elements (Sr, Ba, Ca etc.), have received  a lot of attention due to their fascinating range of physical behavior \cite{ref1} including metal-insulator transition, spin state transition, electron correlations, charge/orbital degree of freedom and colossal magnetoresistance. These exciting physical properties can be attributed to small perturbation due to doping, application of magnetic field, defects, strain etc. such that these materials can serve as a better candidate for spintronics devices and sensors. It has been widely investigated \cite{ref1} that substitution at Mn sites with other transition elements is an efficient way to tune magnetic properties in perovskite oxide materials. Among all transition elements, Co substitution at Mn site has been very important because it has different oxidation states (Co$^{2+}$, Co$^{3+}$ and Co$^{4+}$) and spin states (Low spin, Intermediate Spin and High spin). Substitution at Mn site leads to interaction between the dopant and Mn as well as affects the ratio of  Mn$^{3+}$ and Mn$^{4+}$, which changes the amount of e$_g$ electron in the system. In the present work, we focus on the combined effect of Co and Mn at $B$ sites in La$_{0.7}$Sr$_{0.3}$Mn$_{1-x}$Co$_x$O$_3$ (LSMCO) with ($0 \leq x \leq 1$).

\section{Experimental Details}
LSMCO series ($0 \leq x \leq 1$) were prepared by standard solid state route. All the precursor material were first heated at 800$^o$C and then stoichiometric amount of La$_{2}$O$_{3}$, SrCO$_{3}$, Mn$_{2}$O$_{3}$ and Co$_{3}$O$_{4}$ were mixed and annealed at 1200$^o$C for 22 hours with intermediate grinding. The X-ray diffraction (XRD) pattern was observed using Rigaku diffractometer (Cu-K$_{alpha}$ radiation) at room temperature to find the structure and lattice constants of the LSMCO series. The Rietveld refinement was done to simulate the experimental XRD pattern. The magnetic measurements of the entire series were completed using a Vibrating Sample Magnetometer (VSM) in magnetic field ranging from -5\,T to 5\,T and temperature ranging from 10\,K to 300\,K.  

\section{Results and Discussion}
XRD pattern of LSMCO samples shows that entire series is of single phase with no impurity within its sensitivity, as shown in Fig.~1. The Rietveld refinement of LSMCO series were done with Fullprof software and refinement parameters were given in Table.~1.  The refinement confirms that these samples fall into rhombohedral lattice and space group R$-3$c with lattice parameters, volume and goodness of fit as shown in Table.~1. For LSMCO series, we have kept A site fixed and substitution was done at Mn site. As clear from Table.~1, substituting Co at Mn sites reduces the lattice parameters and hence the volume of unit cell. This is because Co has less ionic radii than Mn.

DC Magnetization M (emu/g) as a function of temperature in Field Cooled (FC) and Zero Field Cooled (ZFC) is shown in Fig.~2 (inset). In FC mode, the magnetization is denoted as $M_{\rm FC}$ and in ZFC mode it is $M_{\rm ZFC}$.  It is seen from Fig.~2 (inset). $M_{\rm FC}$ decreases with increasing temperature, while M$_{\rm ZFC}$ first increases to a maximum value at temperature T$_{\rm g}$ (glass temperature) then decreases with increasing temperature. These two curves meet at a temperature T$_{\rm m}$ for each sample. The value of T$_{\rm g}$ and T$_{\rm m}$ decreases as Co content in the sample increases \cite{ref4}. The behavior of these FC and ZFC curve imply a spin-glass (SG) state arising due to competition between FM \cite{ref2} and AFM states. It has been observed that the maximum value of magnetization decreases as Co content in the sample increases. This signifies that Co content suppresses the long range FM ordering. It is known that La$_{0.7}$Sr$_{0.3}$MnO$_3$ is a FM conductor at room temperature attributed to the double exchange \cite{ref3} (DE) interaction between the Mn$^{3+}$/Mn$^{4+}$ ions. The strength of DE interaction depends on ratio of Mn$^{3+}$/Mn$^{4+}$ ions in the sample. As we substitute Co in place of Mn, the ratio of Mn$^{3+}$ to Mn$^{4+}$ reduces which affects the ferromagnetic properties in the sample.
In FC curve, even a small magnetic field ($0.01$\,T) leads to reorientation of anti-parallel spins to a parallel aligned spin state. On substituting Co close to 1, the ferromagnetic ordering is due to Co ions; the magnetic properties are attributed to the DE interaction between Co$^{3+}$ and Co$^{4+}$ via O$^{2-}$. It has also been observed that substitution of Co leads to reduction in Curie temperature (T$_C$) as shown in Figure.~2.
The M-H hysteresis loops were measured at 10\,K from -2\,T to +2\,T for La$_{0.7}$Sr$_{0.3}$Mn$_{0.90}$Co$_0.10$O$_3$, shown in Fig.~3 and saturation magnetization (M$_{sat}$) and Coercive field (H$_c$) as a function of Co content is shown in Fig.~4. The Co free sample gets saturated near 1\,T but the samples containing Co and Mn do not saturate even at 5\,T (not shown). This is due to competition between FM and AFM interaction which is arising due to mixed Co and Mn state present in the sample. In parent compound the FM state originates due to DE interaction between Mn$^{3+}$-O$^{2-}$-Mn$^{4+}$ while in Co substituted sample the FM state is due to Co$^{2+}$-O$^{2-}$-Mn$^{4+}$ as  Mn$^{3+}$ + Co$^{3+}$ = Mn$^{4+}$ + Co$^{2+}$ and the AFM state is the due to super-exchange interaction of  Co$^{3+}$-O$^{2-}$-Co$^{3+}$, Co$^{2+}$-O$^{2-}$-Co$^{2+}$, Co$^{2+}$-O$^{2-}$-Mn$^{3+}$, Mn$^{3+}$-O$^{2-}$-Mn$^{3+}$, and  Mn$^{4+}$-O$^{2-}$-Mn$^{4+}$. It has been observed from M-H curves, low Co substituted samples have large spontaneous magnetization and as Co content increases, the spontaneous magnetization decreases. This implies  that in Co free samples, half filled cells in Co ions leads to low magnetic properties. The increase in coercive force with increasing Co content in the samples also confirms the suppression of ferromagnetic DE interaction and increase in AFM interaction due to super-exchange.

\section{Conclusion}
In this report, we studied the effect of Co substitution on structural and magnetic properties of La$_{0.7}$Sr$_{0.3}$MnO$_3$ polycrystalline samples. The Rietveld refinement of XRD patterns showed phase purity of the polycrystalline samples. The interplay of FM and AFM interaction upon Co substitution at Mn site in La$_{0.7}$Sr$_{0.3}$MnO$_3$ is evident from magnetic measurements.

\newpage

\begin{table}
\centering
\caption[Structural parameter of La$_{0.7}$Sr$_{0.3}$Mn$_{1-x}$Co$_x$O$_3$ ($0 \leq x \leq 1$) obtained from Fullprof refinement]{Structural parameter of La$_{0.7}$Sr$_{0.3}$Mn$_{1-x}$Co$_x$O$_3$ ($0 \leq x \leq 1$) obtained from Fullprof refinement.}
\begin{tabular}{| c | c | c | c |}
\hline
Co content & Lattice Parameter ($\AA$)  & Volume ($\AA^3$) & $\chi^2$\\
\hline
\hline
x=0.00 & 5.511, 5.511, 13.369 & 351.66 & 1.66\\
\hline
x=0.05 & 5.504, 5.504, 13.362 & 350.55 & 1.54\\
\hline
x=0.10 & 5.508, 5.508, 13.362 & 350.99 & 2.27\\
\hline
x=0.30 & 5.505, 5.505, 13.358 & 350.69 & 1.61\\
\hline
x=0.50 & 5.466, 5.466, 13.470 & 348.60 & 1.68\\
\hline
x=0.70 & 5.463, 5.463, 13.249 & 342.08 & 1.43\\
\hline
x=0.90 & 5.447, 5.447, 13.215 & 339.58 & 1.26\\
\hline
x=0.95 & 5.442, 5.442, 13.201 & 338.52 & 1.61\\
\hline
x=1.00 & 5.396, 5.396, 13.295 & 335.32 & 1.13\\
\hline

\end{tabular}
\end{table}
\newpage
\begin{figure} 
\includegraphics[height=13.0 cm]{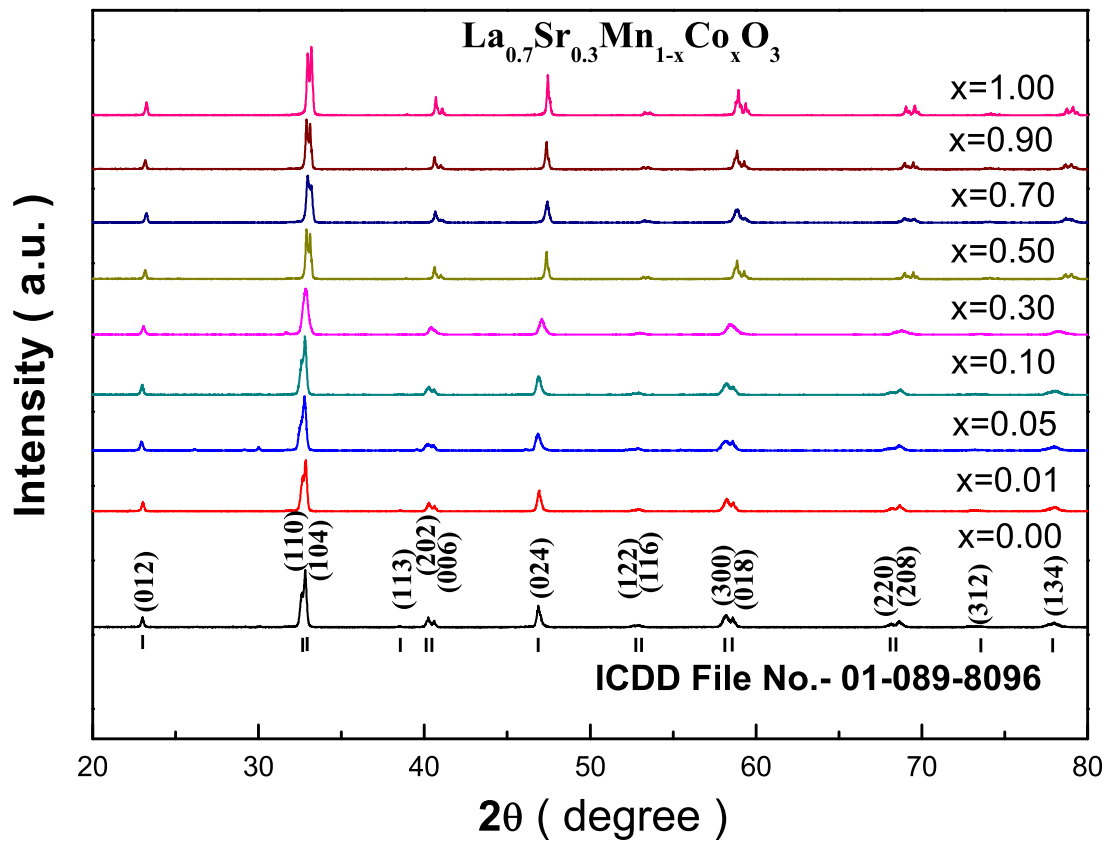}
\caption{(Color online) XRD pattern of polycrystalline La$_{0.7}$Sr$_{0.3}$Mn$_{1-x}$Co$_x$O$_3$ series ($0 \leq x \leq 1$).}
\end{figure}

\begin{figure} 
\includegraphics[height=12.0 cm]{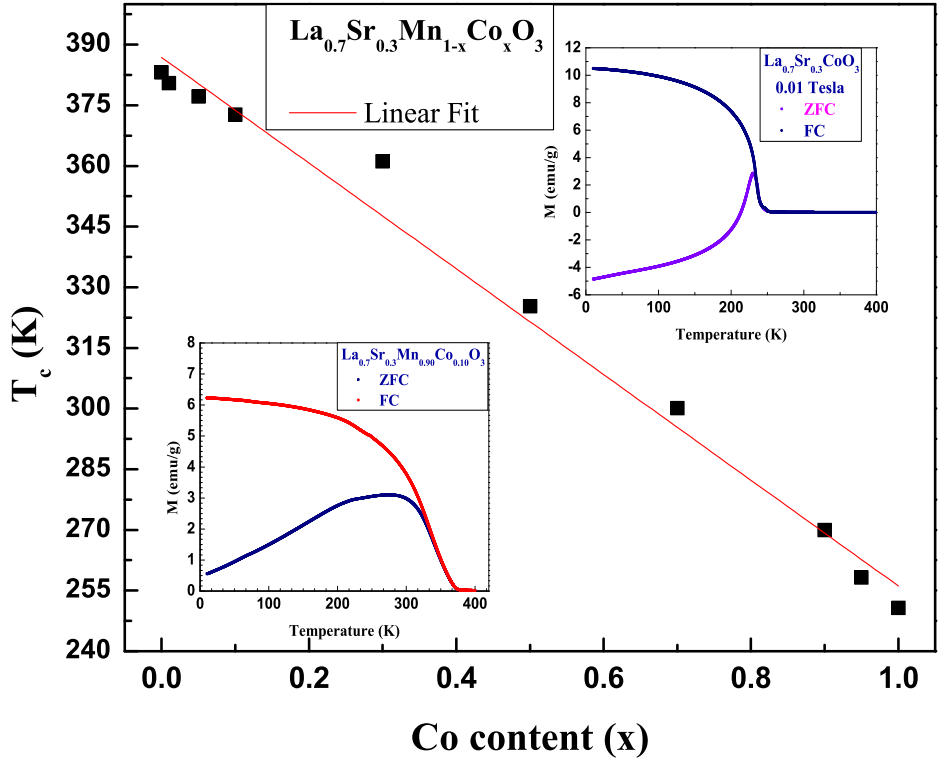}
\caption{(Color online) Curie Temperature (T$_C$)  as a function of Co content in La$_{0.7}$Sr$_{0.3}$Mn$_{1-x}$Co$_x$O$_3$ series. ZFC and FC curve is shown in the inset for LSMC$_{0.1O}$ (lower) and LSCO (upper) samples.}
\end{figure}

\begin{figure} 
\includegraphics[height=12.0 cm]{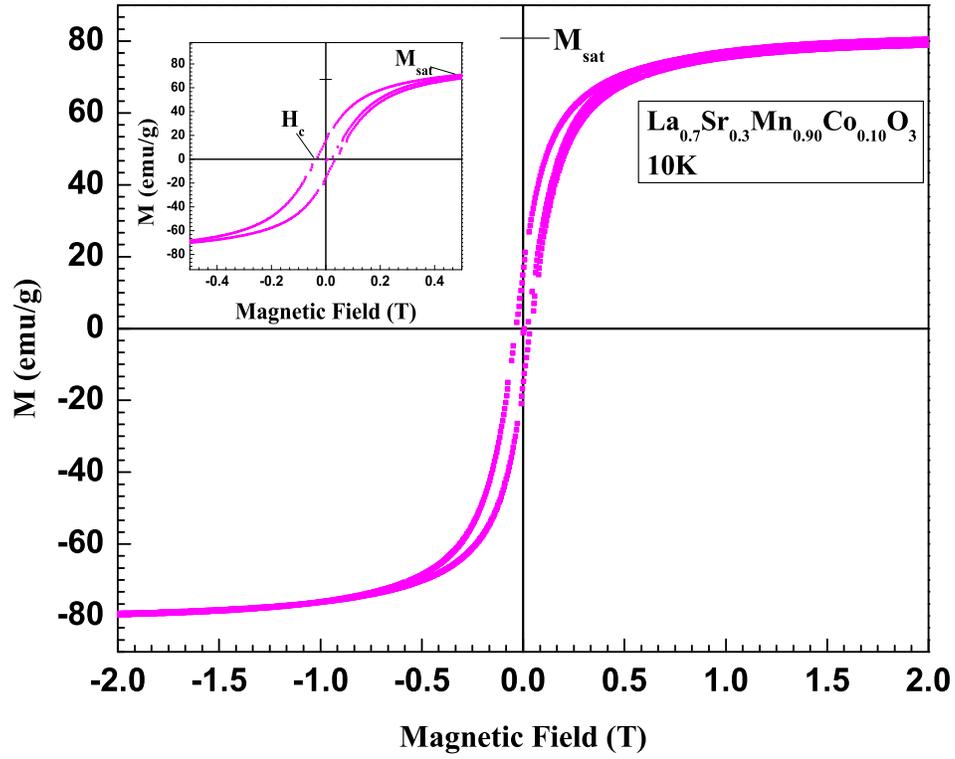}
\caption{(Color online) M-H curve for La$_{0.7}$Sr$_{0.3}$Mn$_{0.90}$Co$_{0.10}$O$_3$ at 10\,K. Inset shows the saturation magnetization (M$_sat$) and coercive field (H$_C$) of La$_{0.7}$Sr$_{0.3}$Mn$_{0.90}$Co$_{0.10}$O$_3$.}
\end{figure}

\begin{figure} 
\includegraphics[height=13.0 cm]{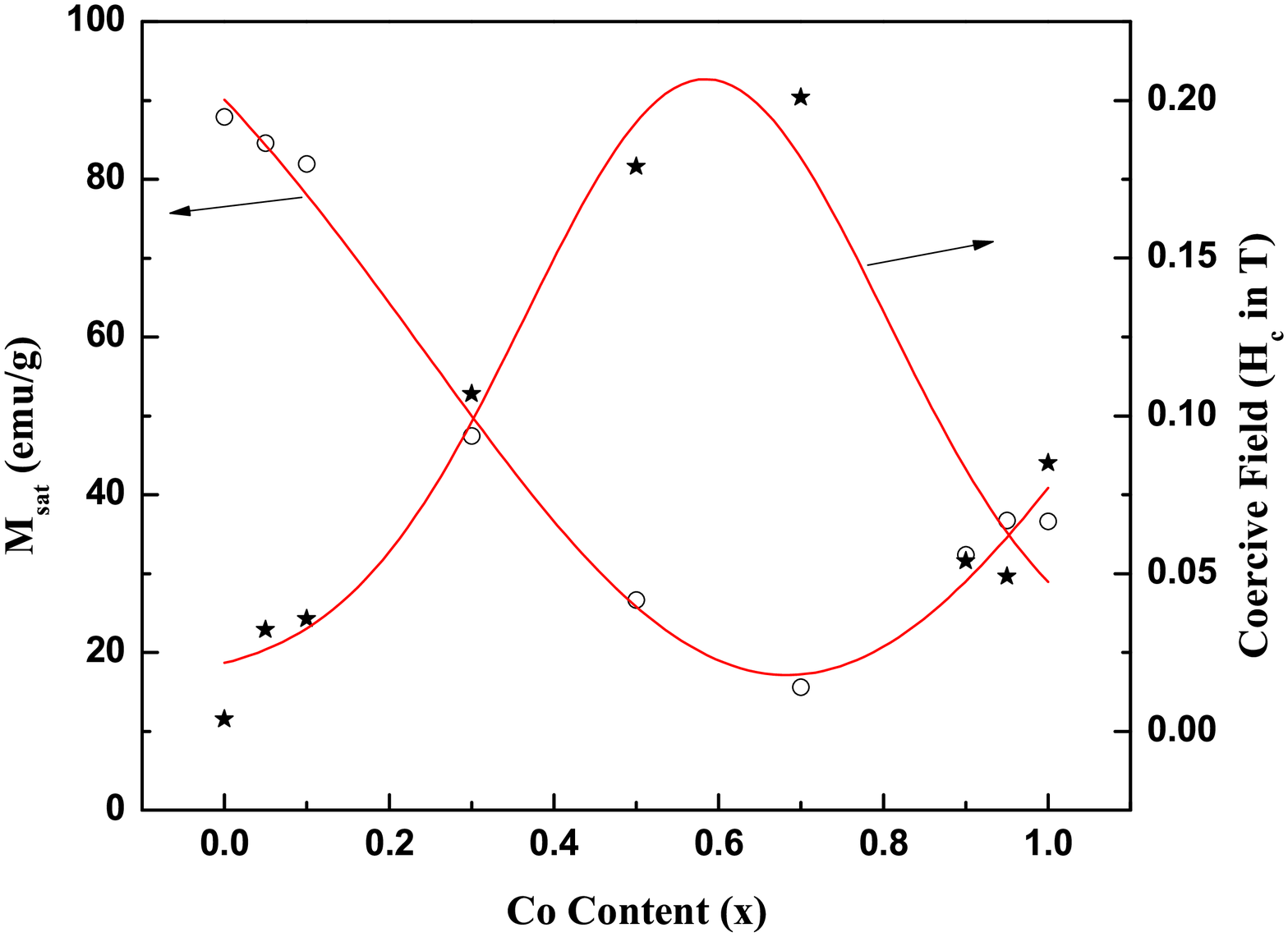}
\caption{(Color online) M$_{sat}$ and H$_C$ as a function of Co content in LSMCO series.}
\end{figure}


\begin{references}

\bibitem{ref1}
M. B. Salamon and M. Jaime, Review of Modern Physics, 73, 583-628, (2001). 
\bibitem{ref2}
A. S. Alexandrov and A. M. Bratkovsky, Journal of Applied Physics, 85, 4349-4351. (1999).
\bibitem{ref3}
P. G. de Gennes, Physical Review, 118, 141-154 (1960). 
\bibitem{ref4}
X. J. Fan et al. J. Phys.: Condens. Matter, 11, 3141-3148 (1999).

\end{references}
\end{document}